\documentclass[longnamesfirst,apjl]{emulateapj}
\usepackage{apjfonts,natbib,epsfig}
\tighten

\newlength{\tskip}
\newlength{\colwidth}

\newcommand{\beq}{\begin{equation}}      
\newcommand{\eeq}{\end{equation}}      
\newcommand{\beqa}{\begin{eqnarray}}      
\newcommand{\eeqa}{\end{eqnarray}}      

\begin{document}      
\title{IR Background Anisotropies in Spitzer GOODS images 
and constraints on first galaxies} 
\author{Asantha Cooray\altaffilmark{1},
Ian Sullivan\altaffilmark{2}, 
 Ranga-Ram Chary\altaffilmark{2},      
James~J.~Bock\altaffilmark{2,3},
Mark Dickinson\altaffilmark{4}, 
Henry C. Ferguson\altaffilmark{5}, 
Brian Keating\altaffilmark{6}, Andrew Lange\altaffilmark{2},
Edward L. Wright\altaffilmark{7}}  
\altaffiltext{1}{Center for Cosmology, Department of Physics and Astronomy, University of California, Irvine, CA 92697. E-mail: acooray@uci.edu}      
\altaffiltext{2}{Division of Physics, Mathematics, and Astronomy, California Institute of Technology, Pasadena, CA 91125}      
\altaffiltext{3}{Jet Propulsion Laboratory, 4800 Oak Grove Drive, Pasadena, CA 91109}      
\altaffiltext{4}{NOAO, 950 N. Cherry Ave., Tucson, AZ 85719}    
\altaffiltext{5}{STSCI, 3700 San Martin Dr., Baltimore, MD 21218}    
\altaffiltext{6}{Department of Physics, University of California, La Jolla, CA}
\altaffiltext{7}{UCLA Astronomy, PO Box 951562, Los Angeles, CA 90095-1562}

\begin{abstract}      
We describe the angular power spectrum of unresolved 3.6 $\mu$m IR light 
in {\it Spitzer} GOODS fields.
The amplitude of the anisotropy spectrum decreases with decreasing flux threshold to which resolved sources are removed from images. 
When all pixels brighter than a Vega magnitude of 24.6 are removed, the amplitude of the power spectrum at arcminute angular scales 
can be described with an  extra component of $z>8$ sources with a IRB contribution around 0.4 nW m$^{-2}$ sr$^{-1}$. 
The shape of the power spectrum, however, is more consistent with that expected for unresolved, faint galaxies at lower redshifts with Vega magnitudes fainter than 23 with a total 3.6 $\mu$m intensity between $0.1$ to $0.8$ nW m$^{-2}$ sr$^{-1}$. We confirm this
assumption by showing that large-scale power decreases rapidly  when the unresolved clustering
spectrum is measured  from a processed HDF-N IRAC image where locations of faint ACS sources with no IR counterparts were also masked. 
Based on resolved counts and unresolved fluctuations, we find that, at most,  about 
7.0 nW m$^{-2}$ sr$^{-1}$ can be ascribed to galaxies.
\end{abstract}      
      
\keywords{large scale structure of universe --- diffuse radiation --- infrared: galaxies}      
      
\section{Introduction}      
The intensity of the cosmic near-infrared background (IRB) is a measure of the total light emitted by stars and galaxies in the      
Universe. While the absolute background has been estimated by space-based      
experiments, such as the  Diffuse Infrared Background Experiment (DIRBE; Hauser \& Dwek 2001) and       
the Infra-Red Telescope in Space (IRTS; Matsumoto et al. 2005), the total measured IRB intensity      
still remains unaccounted for by sources. At 3.6 $\mu$m ($L$-band), 
the total IRB intensity is 12.4 $\pm$ 3.2 nW m$^{-2}$ sr$^{-1}$ (e.g., Wright \& Reese 2000) and resolved sources
in {\it Spitzer} lead to about 5.4 nW m$^{-2}$ sr$^{-1}$
(Fazio et al. 2004) to  $\sim$ 6.0 nW m$^{-2}$ sr$^{-1}$ (Sullivan et al. 2006). 

On the other hand, fluctuation analysis of the {\it Spitzer} images have shown an excess anisotropy, which has been
attributed to first-galaxies containing Population III (Pop-III) stars at redshifts before reionization (Kashlinsky et al. 2005). 
This possibility is motivated by suggestions in the literature that Pop-III stars can explain the
difference between measured and resolved total IRB intensity
(e.g., Santos, Bromm \& Kamionkowski 2002; Salvaterra \& Ferrara 2003; Cooray \& Yoshida 2004). If this were to be the case, then
this high-redshift component is best studied with intensity anisotropies of the IRB since a
high-redshift clustering spectrum differs in shape from low-redshift clustering
(Cooray et al. 2004; Kashlinsky et al. 2004). In a recent paper (Sullivan et al. 2006), we discussed  IR intensity
fluctuations from resolved sources in {\it Spitzer} images down to a Vega mag. of 22.5. 
In Sullivan et al (2006), we also suggested that the anisotropies in the unresolved background (e.g., Kashlinsky et al. 2005)
are more likely be due to unresolved, faint galaxies at redshifts between 1 and 4 and not from a new population of high-redshift
galaxies with an IRB intensity of $\sim$ 1 to 2 nW m$^{-2}$ sr$^{-1}$ in  the $L$-band (Salvaterra et al. 2006; Kashlinsky et al. 2005).

In this paper, we make a new set of clustering measurements in the 
the Great Observatories Origins Deep Survey (GOODS; Dickinson et al. 2003). 
Instead of fluctuations associated with resolved light, we measure the
anisotropy power spectrum in multipole space of the unmasked pixels
after removing resolved sources down to  a certain flux limit. 
With pixels removed down a magnitude level fainter than 24.6,
we find that the background power spectrum is generated by faint galaxies at
redshifts between 1 to 4, though we cannot  establish precisely 
the slope of the faint-end 
number counts below the point source detection level.

The {\it Letter} is organized as following: in the next section, we briefly summarize 
the procedure we used to measure clustering of unresolved IRB light.
In \S~3 we present  our results and discuss implications of our measurements.
Throughout the paper, we refer to Vega magnitudes ($m_{\rm Vega}=0$ is 280.9 Jy). 
For numerical models,
we make use of a flat-$\Lambda$CDM cosmology with parameters $\Omega_m=0.3$,  $h=0.7$, and a normalization to
the matter power spectrum today at 8 h$^{-1}$ Mpc scales $\sigma_8=0.84$.

\begin{figure*}[!t]      
\centerline{\psfig{file=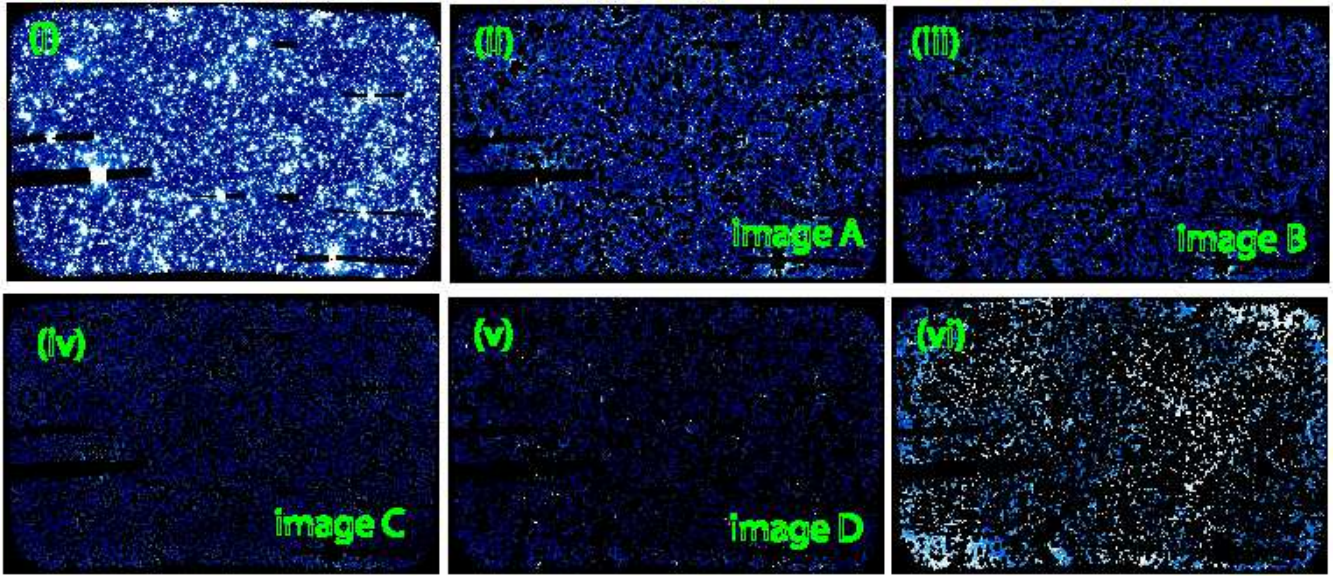,width=7.0in,angle=0}    }
\caption{
Original GOODS HDF-N image (i) and various masked versions of the same region (all panels use the same intensity scale). 
Panels (ii), (iii), (iv), and (v)
correspond to Images A, B, C, and D, respectively (see Table~1). Panel (vi) is
a simulated image of PopIII fluctuations imprinted on
image B, which has an additional mask to describe locations of faint optical Hubble/ACS sources with no IRAC counterparts.}
\label{fig:irb}      
\end{figure*}      

\begin{table}[!t]\small      
\caption{\label{table:data}}    
\begin{center}      
\begin{tabular}{ccccc}      
\hline
Image & Point Source  & 4$\sigma$ & Fraction of & $C_l^{\rm SN}$   \\
      & Removal Limit &  flux/nJy    &   Pixels Used & (nW$^2$ m$^{-4}$ sr$^{-1}$) \\
\hline
A & 19.2 & 105/183 &  55\% & $(1.52/1.45) \times 10^{-11}$\\
B & 20.2 & 72.8/66.4 &  46\% & $(1.00/1.01) \times 10^{-11}$\\
C & All & 41.9/41.9 &  30\% & $(5.97/6.35) \times 10^{-12}$\\
D & B+ACS & 68.6/64.3 & 20\% & $(7.3/8.1) \times 10^{-12}$\\  
\hline      
\end{tabular}\\[12pt]      
\begin{minipage}{3.5in}      
NOTES.---%
Parameters of 3.6 $\mu$m {\it Spitzer} GOODS images used for IRB
anisotropy measurements. Image C removes all sources down to some arbitrary magnitude (see text), while
image D includes an additional mask for pixels in which faint point sources are present 
in Hubble/ACS GOODS catalogs, but no IR counterparts.
The two sets of numbers are for HDF-N/CDF-S, and
``4 $\sigma$ Limit'' is the flux (in nJy) of the brightest remaining pixel after removing pixels in the image at 4 $\sigma$ above
the rms measured from unmasked pixel, with clipping only performed once. 
\end{minipage}       
\end{center}      
\end{table}

\section{Imaging Data of GOODS}      
      
Imaging data of GOODS fields (both CDF-S and HDF-N) with {\it Spitzer} IRAC
were first reduced using a point kernel for drizzling (Fruchter \& Hook 1998).
In both fields, sources were detected and masked out from the images using three different techniques.
One technique was identifying all sources in the SExtractor GOODS IRAC 
catalog (Bertin \& Arnouts 1996) that used Mexican hat convolution kernels to optimize source deblending and 
tuned to push the surface brightness threshold for detection down to the faintest levels. The diameter of the circular
mask for each source was scaled by the source brightness. 
Sources which are brighter than 18.3 mag has a 7.2'' diameter mask. This scaled linearly down to a 4.8'' diameter 
mask for sources at 19.3 mag. The choice of masking diameter is defined by the point spread function which has its 
first Airy bright 
ring at ~2.4'' distance from the center (the pixel scale is 0.6'' in the GOODS IRAC mosaics). All sources fainter than 19.2 mag
were masked with a 4.8'' diameter mask.  We also generated masks from the SExtractor segmentation maps. In this procedure 
every source has a region defined for it in the segmentation map. We grew this region by 2 pixels to ensure that the 
wings of the sources are masked as well (see Table~1). In addition, we also generated an HDF-N image 
which masked sources that were present in the Hubble/ACS catalogs of GOODS fields (Giavalisco et al. 2004) 
but were absent in the IRAC catalog (image D).  These are typically faint, blue galaxies in $B_{\rm 435}$ and $V_{\rm 606}$ ACS
bands and must lie at $z<5$.
Finally, a third procedure was used to mask out all sources (image C). This involved smoothing the 
image by a 3$\times$3 boxcar, identifying a sky level by using a 15$\times$15 boxcar and finding pixels which are more than 2 sigma above the sky level, when the width of the aperture used to mask out the pixels is 3 pixels wide.

In each of the images, when measuring clustering, we additionally
discard pixels with intensities that are 4 $\sigma$ above the mean pixel flux level estimated by averaging all background pixels
that were not masked with the initial source detection. This clipping, done once, allows us to create an image where the brightest
pixel has a flux of 183 nJy (image A) to 42 nJy (image C).  Note that given the width of the IRAC PSF, a brightest pixel of 42 nJy 
would correspond to a source with a Vega magnitude of about 22.1 (0.4 $\mu$Jy).
The 1$\sigma$ flux uncertainty for isolated point sources over most of GOODS 3.6 $\mu$m area is 22 nJy and corresponds to a formal
signal-to-noise of 5 point-source detection limit of 23.5 in Vega magnitudes. From simulations, 
the 50\% completeness limit of IRAC Sextractor catalogs is about 22.5.

When measuring clustering of the empty pixels or the unresolved IR light,
we take Fourier transforms (FT) of the images with masked pixels filled with white noise corresponding to the rms level of the image.
In Thompson et al. (2006), the masked pixels were set to an intensity unit of zero. We tested both options and
found that setting the masked pixels to a zero intensity leads to a power spectrum that is larger in amplitude
by 25\% at angular scales corresponding to typical distances between large masked areas. This effect is likely negligible for the Thompson et al. (2006) analysis, due to the much smaller Airy disk of the NICMOS instrument. 

In general, each image is described by an intensity distribution at each of the pixels $I(\vec{\theta})$ where $\vec{\theta}$
is a two-dimensional angular vector on the sky. We take the Fourier Transform of 
the image $I_\ell(\vec{\ell}) \propto \int I(\vec{\theta}) e^{-i\vec{\ell} \cdot \vec{\theta}} d^2\theta$, where $\ell$ is the wave number. 
The angular power spectrum is constructed as $C_\ell=|I_\ell|^2$ (and plotted in Fig.~2). In practice,  note that we take discrete transforms
and the procedure is  analogous to the process described in
Appendix A of Thompson et al. (2006).  We believe a similar procedure
was used in Kashlinsky et al. (2005), but we have not been able to implement their  exact technique used for point source detection and
masking.   In Fig.~2,  we use bootstrap sampling  to estimate the error bars of the binned power spectrum estimates.
While we plot the total angular power spectrum, at small angular scales the clustering is dominated by a combination of a
finite-density source shot-noise, 
$C_l^{\rm SN} =\int S^2 (dn/dS) dS$ below  the cut-off flux $S_{\rm cut}$, and instrumental noise. 
Here, we use the shape of the power-spectrum at large  $\ell$ to measure the shot-noise directly from clustering spectra
and compare with expectations given models of counts at the faint-end. 

To address the possibility that the clustering spectra may be
affected by residual flux from bright sources that are not perfectly masked out,
we also cross-correlated images at different magnitude cuts with catalogs of the extracted sources. 
We show the cross-correlation power between  detected sources fainter than 21st magnitude and image C in Fig.~2.
This spectrum is below the power measured in images alone and allows us to safely assume that
the large-scale clustering pattern is not simply a result of residual fluxes next to brighter sources. 
Finally, to estimate any systematics common to the two fields, we cross-correlated C images of HDF-N and CDF-S fields
with imaging arrays centered on a common coordinate system.  As shown in Fig.~2, 
this cross-correlation spectrum is well below signal spectra in each of the fields.

\begin{figure}[!t]      
\centerline{\psfig{file=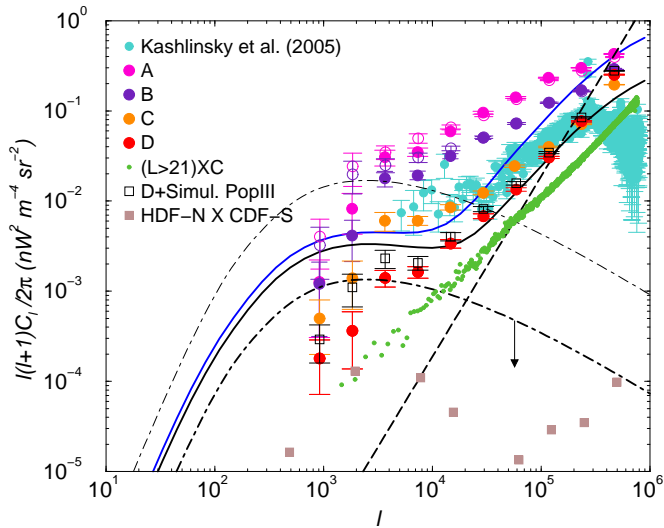,width=2.8in,angle=-90}}
\caption{
Clustering of the unresolved IRB light in GOODS HDF-N (filled symbols) and CDF-S (open symbols for images A and B) fields as a function of the
magnitude cut in $L$-band for resolved source detection. For reference, we also show the angular power spectrum of
anisotropies from Kashlinsky et al. (2005) in light cyan color. With decreasing flux cut the amplitude of clustering decreases,
with the minimum reached in image D.
The smaller points in green are the
cross-correlation between an image made with sources in resolved source catalogs with $m(3.6\mu {\rm m}) > 21$
and the HDF-N image C.
This sets a lower limit on the confusion that arises from residual flux form wings of the brighter resolved sources that are not
properly removed from the images. Finally, smaller points in brown are the cross-correlation between C images of HDF-N and CDF-S.
This sets the lower limit for any common systematics from IRAC detectors.
The lines show expected levels of angular clustering in the IRB:
the top and bottom dot-dashed lines are contributions from a population of Pop-III containing galaxies  at  $z>8$ with an intensity contribution to 
the IRB of 1.5 nW m$^{-2}$ sr$^{-1}$ and 0.4nW m$^{-2}$ sr$^{-1}$ at 3.6 $\mu$m, respectively (see Cooray et al. 2004 for details). 
The solid lines show the clustering spectrum of resolved and unresolved galaxies based on models for number count in Fig.~3
using the conditional luminosity function descriptions in Sullivan et al. (2006). 
The dashed line that scales as $l^2$ with increasing multipole is the measured shot-noise in image C.
}
\end{figure}

\begin{figure}[!t]      
\centerline{\psfig{file=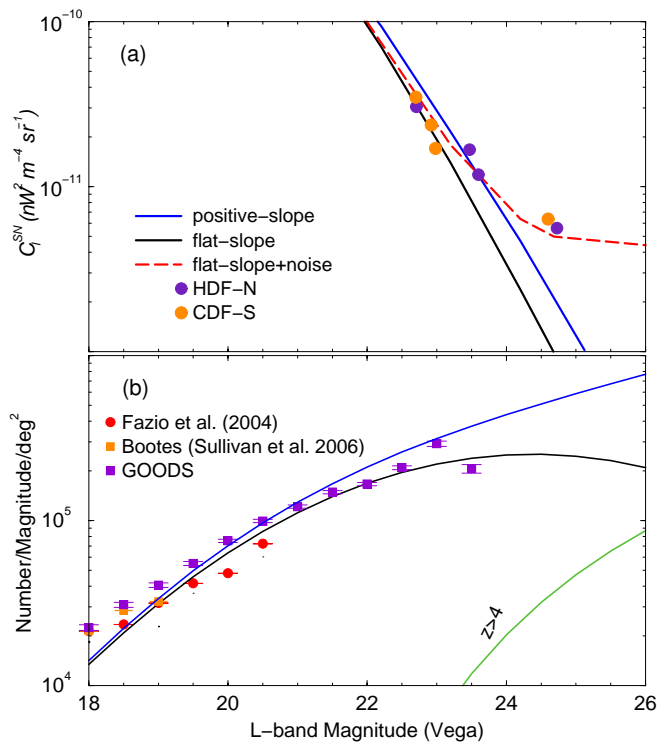,width=3.4in,angle=0}}
\caption{
(a): Shot-noise level of clustering as a function of source removal magnitude (solid lines correspond to counts in panel b)
while  filled (HDF-N) and open (CDF-S) symbols show the measured shot-noise in the 4$\sigma$ clipped
images (see Table~1). In addition to the finite density of sources, our measured shot-noise also includes a noise floor
related to detector noise, which is a constant as a function of the magnitude cut.
(b): Measured source counts and models based on conditional luminosity functions (from Sullivan et al. 2006).
The shot-noise level in the images with sources masked out agrees with the expectation from counts and there is no 
evidence for a new population of sources at magnitudes fainter than point source detection.
}
\label{fig:sn}      
\end{figure}

\section{Results \& Discussion}      

In Fig.~2 we compare our measurements with the anisotropy spectrum from
Kashlinsky et al. (2005).  At the faintest magnitude where sources in the GOODS images are removed (image C), our fluctuation spectrum agrees
with their measurements, though clustering in GOODS images probe a factor of two larger angular scales.
At the largest angular scales probed by our measurements, the clustering power decreases relative to fluctuations
presented in Kashlinsky et al. (2005). Such a decrease is inconsistent with
the description that clustering of IR light is associated with $z>8$ first galaxies with a 3.6 $\mu$m IRB intensity between
1  to 2 nW m$^{-2}$ sr$^{-1}$  (dot-dashed line in Fig.~2; e.g., Salvaterra et al. 2006). 

To test the origin of fluctuations, beyond the analysis in Kashlinsky et al. (2005),
we produced images where we additionally masked pixels in which faint Hubble/ACS sources involving
mostly blue galaxies at $z<5$ (Giavalisco et al. 2004) are present with no IR counterparts.
The spectrum of the residual background with this image (image D) 
has the lowest amplitude  at arcminute angular scales and support
the hypothesis that the excess IR clustering is due to faint galaxies.
If we simply scale that power spectrum down
to match the amplitude of the clustering spectrum at $\ell$ around 5000, the implied intensity of
IR light from such a $z>8$ component is about 0.4 nW m$^{-2}$ sr$^{-1}$. There is a large uncertainty
in modeling the expected clustering of PopIII fluctuations related to gas cooling and
star-formation efficiency among others (see, Cooray et al. 2004). To compare with previous
interpretations, the above intensity is determined with parameters similar to those in Salvaterra et al. (2006).

Since there is a possibility
that PopIII fluctuations may be removed during the masking and point source removal,
we simulated PopIII backgrounds with power spectra given by dot-dashed lines in Fig.~2. 
We added these simulated maps (see, Fig.~1 panel vi) to GOODS images and repeated the same point source detection and
masking procedure 
The measured clustering spectrum of this image recovers the original input spectrum  of PopIII fluctuations within 10\%.
The open squares in Fig.~2 show the expected spectrum if GOODS HDF-N contained a PopIII background with an intensity of
of 0.4 nW m$^{-2}$ sr$^{-1}$ corresponding to the lower dot-dashed line in Fig.~2. At multipoles of
10$^3$ to 10$^4$, the power spectrum estimates using the combined resolved-source and faint ACS-source mask is a
 factor of 2 higher than
measured in the corresponding image D. While a careful analysis could further lower the maximum intensity of allowed
PopIII fluctuations, we can safely state that our analysis indicates that the PopIII background in IR images at 3.6 $\mu$m is 
well below the previous suggested level of 1 to 2 nW m$^{-2}$ sr$^{-1}$.

As shown in Fig.~2, the clustering spectra are more consistent with what is expected for faint galaxies. 
We model expected clustering using the same halo model described in Sullivan et al. (2006) for galaxies
with $m(3.6 \mu {\rm m}) >  23$. At the bright-end, these models
are normalized to reproduce  3.6 $\mu$m luminosity functions from the SWIRE survey  (Babbedge et al. 2006).
Since the faint-end shape of the LF is still uncertain, that uncertainty leads to a 
difference in the expected counts at the faint end
as shown in Fig.~3(b), though even with increasing counts down to fainter magnitudes, these
faint galaxies are still restricted mostly to redshifts below 4 ($4 < z < 8$ counts are shown with  the green line in Fig.~3b).
While the measured power spectra are consistent  with clustering of these faint galaxies, 
our results do not allow us to conclusively state if
the number counts flatten or continue to increase with a large positive slope at fluxes fainter than point source detection.
At magnitudes fainter than 23, these counts lead to an additional contribution to IR light between 0.1  and 0.8 nW m$^{-2}$ sr$^{-1}$, for flat and
positive-slope counts, respectively.  Note that the clustering calculations in Fig.~2 are simply an extrapolation from halo models in
Sullivan et al. (2006). We have not attempted to vary any parameters in our model to obtain a better fit.
Given the agreement in Fig.~2 between predictions and measurements, it is clear that one does not need to invoke a new
population of high-z galaxies to explain unresolved IR light clustering. 

We can also explore the number counts of these galaxies with the shot-noise amplitude.
In Fig.~3(a) we compare the measured shot-noise level from each of the images
with the same expected under completeness corrected counts shown in the right panel. There is a general agreement between expected and measured shot-noise level, though again we cannot distinguish between different slopes for counts
precisely. 
With improved measurements, it may eventually be possible for us to address which is the exact slope of the faint-end number counts of IR galaxies.
We leave this work for a later study. Earlier estimates (Fazio et al. 2004; Sullivan et al. 2006) showed that down to a Vega magnitude of about 23, counts lead to an intensity of 6.0 nW m$^{-2}$ sr$^{-1}$  at 3.6$\mu$m.
With the counts increasing with the positive-slope at the faint end, we find that, 
at most, about 7.0 nW m$^{-2}$ sr$^{-1}$ of 3.6 $\mu$m IRB intensity comes from known  galaxy populations. 

Our results do not completely rule out primordial $z > 8$ galaxies with Pop-III stars in the IR background. 
They may still be present with an intensity contribution below 0.4 nW m$^{-2}$ sr$^{-1}$. If they are to explain a
larger intensity, then this background must be extremely smooth at angular scales below 10 arcminutes and, for some reason,
the clustering spectrum must peak at multipoles lower than $10^3$.
Moreover, 
$z >8$ galaxies are expected to be dominating at lower wavelengths than 3.6 $\mu$m. Recent clustering measurements by Thompson et al.
(2006) in the NICMOS Deep Field argue against a large contribution to the IR light at 1.6 $\mu$m. In the near future, the Cosmic Infrared
Background Experiment (CIBER; Bock et al. 2006; Cooray et al. 2004) will measure zodiacal light, total background intensity,
and fluctuations of IR light over 4 square degrees with rocket-borne imagers and spectrometers. It should provide
the conclusive answer on whether a residual in zodiacal light modeling is responsible for the large unexplained intensity of
IR light at these wavelengths.

To briefly summarize our results and discussion,
the  clustering power spectrum of unresolved light in {\it Spitzer} GOODS images are consistent with 
faint galaxies  below the resolved counts and  predominantly at redshifts below 5. These faint galaxies are expected from
extrapolating measured IR LFs, but  remain unresolved even in deep
{\it Spitzer} images though, for most of them, optical counterparts may exist in Hubble/ACS images. 
They are not a significant contributor to the total IRB light, and 
the 3.6 $\mu$m total intensity from resolved sources down to a Vega magnitude of 23 and those needed to
 explain unresolved fluctuations is about 7.0 nW m$^{-2}$ sr$^{-1}$.

{\it Acknowledgments:}       
This work is based on  observations made with the {\it Spitzer Space Telescope}, 
which is operated by the Jet Propulsion Laboratory, California
Institute of Technology, under NASA contract 1407.      
This research was funded by NASA APRA at JPL, Caltech, and UC Irvine.
We thank A. Kashlinsky, R. Arendt, J. Mather, and H. Moseley for communicating their recent results prior to publication
and for useful discussions. Images and data related to this study  are available from http://www.cooray.org .

\end{document}